\documentclass[%
 reprint,twocolumn,floatfix,
 amsmath,amssymb,
 aps,
 prl,
 showpacs,
 superscriptaddress,
 longbibliography,
 floatfix
]{revtex4-2}

\usepackage{graphicx}
\usepackage{dcolumn}
\usepackage{bm}
\usepackage{color}
\usepackage{comment}
\usepackage{url}

\newcommand*{\rr}{\textbf{r}} 
\newcommand{\figurewidth}{0.9\columnwidth}

\begin{document}

\title{Charge-regulation effects in nanoparticle self-assembly}

\author{Tine Curk}
\email{curk@northwestern.edu}
\affiliation{Department of Materials Science \& Engineering,
  Northwestern University, Evanston, Illinois 60208, USA}

\author{Erik Luijten}
\email{luijten@northwestern.edu}
\affiliation{Department of Materials Science \& Engineering,
  Northwestern University, Evanston, Illinois 60208, USA}
\affiliation{Departments of Engineering
  Sciences \& Applied Mathematics, Chemistry, and Physics \& Astronomy,
  Northwestern University, Evanston, Illinois 60208, USA}

\date{\today}

\begin{abstract}
  Nanoparticles in solution acquire charge through dissociation or association
  of surface groups. Thus, a proper description of their electrostatic
  interactions requires the use of charge-regulating boundary conditions
  rather than the commonly employed constant-charge approximation. We
  implement a hybrid Monte Carlo/Molecular Dynamics scheme that dynamically
  adjusts the charges of individual surface groups of objects while evolving
  their trajectories. Charge-regulation effects are shown to
  \emph{qualitatively} change self-assembled structures due to global charge
  redistribution, stabilizing asymmetric constructs. We delineate under which
  conditions the conventional constant-charge approximation may be employed
  and clarify the interplay between charge regulation and dielectric
  polarization.
\end{abstract}


\maketitle

Most solvated materials and biomolecules acquire nonzero charge due to
dissociation or association of charged surface groups. For example, proteins,
DNA, and silica nanoparticles attain their charge via proton
dissociation~\cite{lund13,markovich16}.  The magnitude of the surface charge
is determined by the solution conditions, such as pH, but also by the presence
of other charged entities in the vicinity, via \emph{charge
  regulation}~(CR)\@.  Although this phenomenon has been theoretically studied
since the seminal work by Kirkwood in the 1950s~\cite{kirkwood52}, the
assumption that all objects carry a \emph{constant} charge is still widely
employed.  This is especially striking since the surface charge distribution
plays a key role in colloidal assembly and macromolecular structure formation,
both in materials science~\cite{Boles2016} and in biomolecular systems, such
as protein binding and condensation~\cite{Zhou2018,deSilva09,lund13}.  The
situation is exacerbated by the fact that little is known about the many-body
effects of CR on electrostatic aggregation.

Traditionally, the Poisson equation is solved using the constant-charge (CC)
or the constant-potential (CP) boundary condition.  The CR boundary condition
yields a solution that falls between these two limiting
cases~\cite{ninham71,podgornik18}. Charge-regulation effects have been shown
to change polyelectrolyte~\cite{joanny90} and polymer
brush~\cite{szleifer07,tagliazucchi10} phase behavior and enhance
protein--protein interactions~\cite{deSilva09,lund13}.  Poisson--Boltzmann
theory has elucidated the strong dependence of surface charge on pH, but
is limited to the weak coupling regime and to static and simple geometries
such as flat surfaces~\cite{markovich16}, a spherical particle~\cite{levin19},
or a pair of particles~\cite{popa10,trefalt16}.

Particle-based simulations avoid the approximations inherent to a mean-field
approach. However, whereas molecular dynamics (MD) and Monte Carlo (MC)
simulations of solvated systems with explicit charges are standard, acid--base
dissociation is rarely taken into account. Notable exceptions include hybrid
techniques for atomistic simulations in a constant-pH
ensemble~\cite{chen14,huang16,radak17} and MC investigations of
polyelectrolytes~\cite{kosovan19,murmiliuk18} or planar
surfaces~\cite{madurga11,barr11}. Computational cost has limited these studies
to relatively small systems, so that many-body effects of CR have not been
investigated and its consequences for the self-assembly of charged objects are
largely unknown.

Charge regulation is particularly relevant for aggregation owing to the
relation between the charge distribution and the structure of the aggregate,
which requires self-consistent solution of the problem. Here, we assess the
effects of CR by investigating a fully dynamic system of up to 100 objects
with more than 30,000 explicit, dissociable sites. By implementing an
efficient and parallelizable hybrid MD--MC scheme we examine how aggregation
and self-assembly are affected by CR\@. In addition, we combine our scheme
with the Iterative Dielectric Solver (IDS)~\cite{barros14a,wu18a}, a
boundary-element method, to explore how dielectric polarization, another
intrinsically many-body problem~\cite{ziwei20}, affects CR.

We consider spherical particles with a fixed density of surface-attached
dissociable groups.   Each group, e.g., a weak acid, can be
neutral or charged with a unit charge~$q_{0}$.  The
probability~$\alpha_i$ that a group~$i$ is charged depends on the equilibrium
constant~$\mathrm{pK}_i$ and the chemical potential of the dissociated
ion~$\mu$,
but also on the local electrostatic potential~$\psi(\rr_i)$ at the
position~$\rr_i$ of the group~\cite{ninham71,podgornik18,kosovan19},
\begin{equation}
  \frac{\alpha_i}{1-\alpha_i} = 10^{-\mathrm{pK}_i} e^{-\beta \mu \pm \beta \psi(\rr_i) q_0}\;,
  \label{eq:alpha}
\end{equation}
where $\beta \equiv 1/(k_{\mathrm{B}}T)$ is the inverse temperature and the
$\pm$ applies to negatively (acid) and positively (base) charged groups,
respectively.  Note that
$\Delta \mathrm{pK}_i = \mathrm{pK}_i + \beta \mu \log_{10}(e)$ is independent
of the choice of units.  For acid dissociation in an aqueous solution
$\mu = -\mathrm{pH}\, k_{\mathrm{B}}T\ln(10)$.  Equation~\eqref{eq:alpha}
applies to every dissociable group in the system, so that the full set
$\{\alpha_i\}$ determines the surface charge density
$\sigma(\rr) = \mp \sum_{i} \alpha_i q_0 \delta(\rr_i-\rr)$.  Thereby, these
equations provide a self-consistent boundary condition for the Poisson
equation,
$\nabla \cdot [\varepsilon(\bf r) \nabla \psi(\textbf{r})] =
-\sigma(\textbf{r})$, with $\varepsilon(\bf r)$ the local permittivity.

We apply this scheme to objects immersed in a monovalent electrolyte,
represented via the primitive model, with chemical
potential~$\mathrm{pI}$~\footnote{All concentration-based quantities, e.g.,
  pI, are expressed in units of $l_{\mathrm{B}}^{-3}$}.  Our hybrid MD--MC
method works as follows. The system configuration evolves via conventional MD
using the velocity-Verlet algorithm, with parameters and potentials described
in the Supplemental Material~\cite{SM}.\nocite{wales06}
After every $n_{\mathrm{MD}}$ time
steps, $n_{\mathrm{MC}}$ MC steps are performed, where each step samples the
charging state of a dissociable group, Eq.~\eqref{eq:alpha}, or
inserts/deletes salt ions. To obtain realistic dynamics, the relative
frequency of the MD and MC steps should match the proper ratio between the
particle diffusion and the dissociation rate. Here, however, we focus only on
thermodynamic properties and equilibrium structures, so that the choice of
$n_{\mathrm{MD}}$ and $n_{\mathrm{MC}}$ is guided by efficiency
considerations. As both types of steps have the same computational complexity,
an efficient convergence to equilibrium is obtained by setting
$n_{\mathrm{MD}}=n_{\mathrm{MC}}=1/\delta t$, with $\delta t$ the MD time
step.

Unlike the reaction-ensemble method or the constant-pH ensemble method, which
are restricted to a specific range of pH values and salt
concentrations~\cite{kosovan19}, this scheme consistently implements both salt
ion insertion and solvent dissociation via~$\mathrm{pK_s}$ and is thus valid
for any salt concentration or pH\@.  In addition, within the primitive model
we treat the dissociated charges and the monovalent salt ions as equivalent,
which increases the performance of our MC scheme compared to existing
implementations~\cite{landsgesell20}, cf.\ Supplemental Material.

\begin{figure}
\centering
\includegraphics[width=\figurewidth]{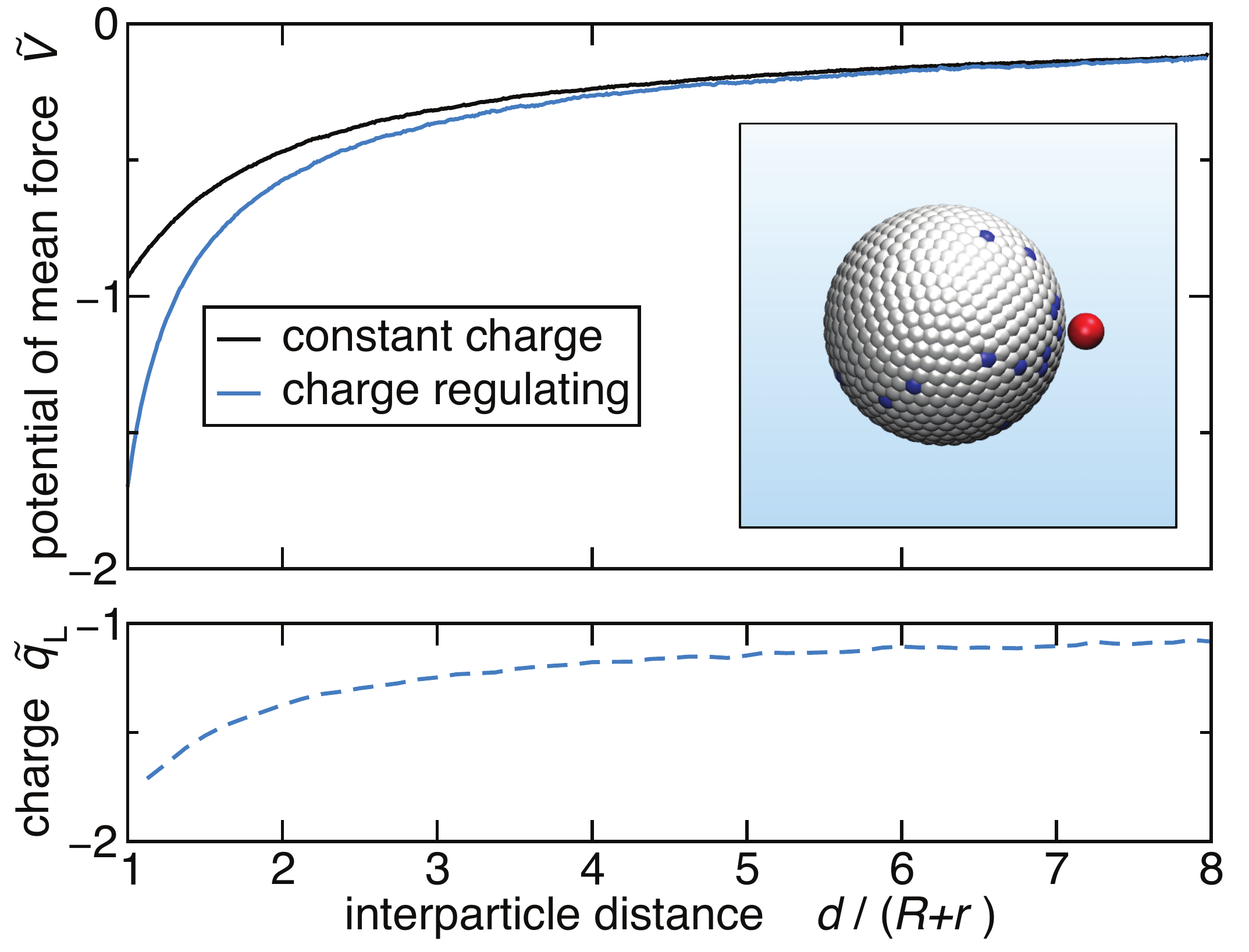}
\caption{Effect of charge regulation on pairwise interactions. Top: Potential
  of mean force~$V$ between a large particle with radius~$R$ and a small
  particle (red sphere) with constant charge~$q_{\mathrm{s}}$ and radius~$r$,
  normalized by the coupling strength~$\lambda$ ($\tilde{V}=V/\lambda$).
  Charge regulation is realized through dissociable sites that can be either
  neutral or charged (white and blue, respectively, in the inset) and results
  in a significantly enhanced attraction (blue line) compared to a constant
  charge $q_{\mathrm{L}}=-q_{\mathrm{s}}$ on the large particle (black line).
  Bottom: Corresponding (normalized) total charge on the larger particle,
  $\tilde{q}_{\mathrm{L}}=q_{\mathrm{L}}/q_{\mathrm{s}}$ (dashed blue line).
  The parameters employed here correspond to a 6-nm silica particle in
  deionized water at $\mathrm{pH}=7$~\cite{trefalt16}; $n_{\mathrm{ss}}=792$
  ($\sigma_{\mathrm{ss}}\approx8\,{\mathrm{nm}}^{-2}$),
  $\Delta \mathrm{pK}=-0.5,$ $q_{\mathrm{s}}=12q_0$, $R=4l_{\mathrm{B}}$,
  $2r=l_{\mathrm{B}}$, $\mathrm{pI}=6.7$, $\lambda=32k_{\mathrm B}T$.}
\label{fig:pmf}
\end{figure}
 
We begin by exploring the influence of CR on a spherical particle covered with
$n_{\mathrm{ss}}=792$ surface sites (Fig.~\ref{fig:pmf}, inset).  To highlight
effects of CR on the electrostatic interactions, we initially disregard
polarization effects as well as London dispersion forces and evaluate the
average charge~$\langle q_{\mathrm{L}} \rangle$ on this particle. Evidently,
this charge depends on the dissociation constant and the solution conditions;
in the Supplemental Material we examine pH dependence and provide a comparison
to Debye--H\"uckel theory. Here, however, we are interested in a more subtle
effect: How does the charge, and thereby interactions, depend on the presence
of other charged entities?  We add a small particle with \emph{constant}
charge equal in magnitude to the charge on the isolated large particle,
$q_{\mathrm{s}} = -\langle q_{\mathrm{L}} \rangle_{d\to\infty}$, and apply the
metadynamics method~\cite{colvars13} to calculate the potential of mean force
(PMF) between the two particles, normalized by the magnitude of the Coulomb
energy at contact $\lambda=q_{\mathrm{s}}^2/[4\pi \varepsilon (R+r)]$ under CC
conditions (Fig.~\ref{fig:pmf}).

Charge regulation results in a nearly twofold increase in the interaction
strength at contact compared to the CC approximation. This arises due to
redistribution of charges on the sphere---an effect similar to polarization of
conducting objects---and due to the change in the total
charge~$q_{\mathrm{L}}$ on the large particle, which depends on the proximity
of the point charge~$q_{\mathrm{s}}$. In the absence of CR effects, an
equivalent \emph{conductive} sphere would yield an increase in the interaction
strength by a factor 1.6~\cite{ziwei20}.  At higher ionic strengths the
electrostatic interactions are screened and weakened. Crucially, however, in
the presence of CR the interaction at contact remains about twice stronger
than the corresponding interaction under CC conditions, even at physiological
salt concentration conditions (see Supplementary Material).

A central challenge in the rational design of materials is the prediction of
structure. Our findings for a particle pair indicate that CR may significantly
affect aggregation. Moreover, dielectric polarization has been shown to induce
large-scale changes to self-assembled structures through local redistribution
of charge \emph{within} particles~\cite{barros14b,ziwei20}. Since charge
regulation allows \emph{global} redistribution of charge, we may expect it to
be an even more powerful factor. Thus, we turn to many-body effects and
self-assembly of multiple particles. Binary mixtures of size-asymmetric
particles give rise to a plethora of self-assembled
structures~\cite{leunissen05,demirors15}. We focus on a prototypical system of
spherical particles with size ratio 1:7, motivated by the observation that CR
appreciably changes the pair interaction when a small particle is positioned 
at approximately~$R/7$ from the surface, or at a center-to-center 
distance~$d \approx 8R/7$ (Fig.~\ref{fig:pmf}).

\begin{figure}
\centering
  \includegraphics[width=\figurewidth]{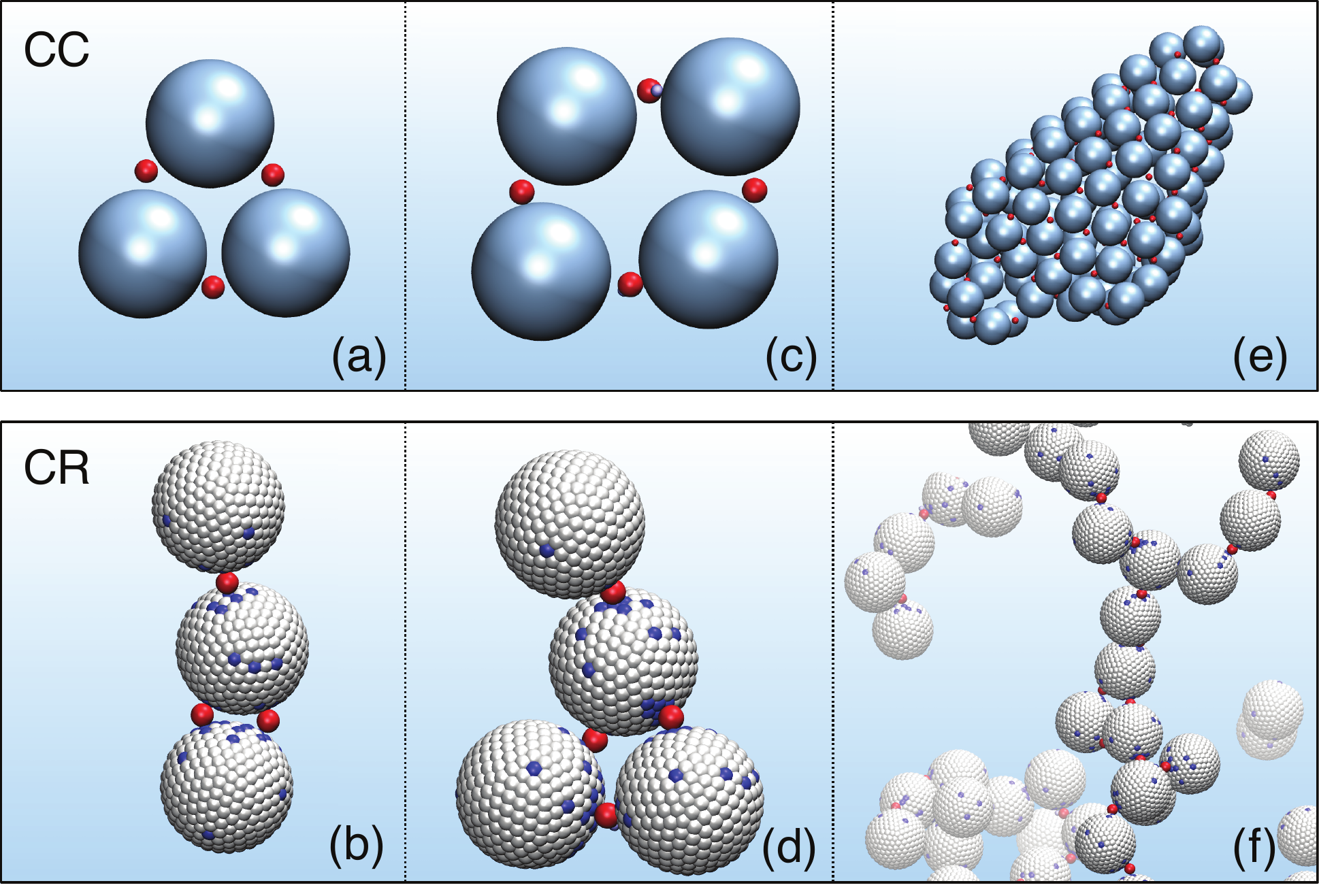}
  \caption{Self-assembly of binary aggregates. Small particles (red) carry a
    constant charge~$q_{\mathrm{s}}=16q_0$ while large particles either have a
    constant charge $q_{\mathrm{L}}=-q_{\mathrm{s}}$ (CC, blue spheres in
    panels a,c,e) or are charge-regulating (CR, spheres with neutral (white)
    or charged (blue) surface sites, panels b,d,f). In the CR case,
    $\Delta \mathrm{pK}=-1.5$, which results in neutral structures,
    $\langle q_{\mathrm{L}} \rangle \approx-q_{\mathrm{s}}$. Whereas CC
    conditions give rise to compact structures, CR leads to anisotropic and
    open assemblies. This observation persists with increasing particle
    number: $N_{\mathrm{p}}=3$, $4$, and $100$ large and small particles in
    panels (a,b), (c,d), and (e,f), respectively, at ion concentration
    $\mathrm{pI}=5.7$.  
    The CR images are instantaneous realizations; the charge distribution on
    the large particles is continuously fluctuating.}
\label{fig:manybody}
\end{figure}

To understand how CR can modulate the local arrangements of building blocks in
a binary mixture, we start with a small system of three large and three small
particles, at concentration $\rho=0.0002 R^{-3}$ and coupling strength
$\lambda = 64 k_{\rm B}T$. We focus on charge-neutral systems to clearly
decouple CR from other possible effects, such as the formation of extended
structures due to nonzero net charge of the aggregate.  In the conventional
(CC) case, the particles form a compact, symmetric aggregate
(Fig.~\ref{fig:manybody}a).  However, when the charge is no longer kept fixed
and identical on each large particle, the CR simulation shows an extended
conformation (Fig.~\ref{fig:manybody}b).  Interestingly, this is accompanied
by symmetry breaking: The average net charge of the three large particles in
Fig.~\ref{fig:manybody}b is
$\bar{q}_{\mathrm{L}} = \{9.8q_0, 22.3q_0, 16.0q_0\}$ for the top, middle and
bottom particles, respectively, indicating that CR stabilizes asymmetric,
heterogeneously charged structures through a \emph{global} redistribution of
charge.

A similar symmetry-breaking transition has recently been reported for CR of
membrane stacks~\cite{majee19}.  A set of four large and four small particles
shows the same trend, forming a symmetric (square-like) structure under CC
conditions (Fig.~\ref{fig:manybody}c), but an asymmetric structure under CR
(Fig.~\ref{fig:manybody}d). This charge redistribution due to CR persists for
larger systems and gives rise to much more extended structures than found in
CC self-assembly. As noted, the CR-induced enhancement of pairwise
interactions continues to hold at higher ionic strength. The same is true for
the asymmetry imparted by CR (see Supplemental Material).

We illustrate this in a system of 100 large and 100 small particles at a
concentration $\rho=0.0043R^{-3}$ (lateral system size $L=28.5 R$).  The
structures are characterized by the local coordination number~$z$, which
measures the number of small particles within a distance~$d_{\mathrm{n}}=R+3r$
(the first minimum of the radial distribution function) from each large
particle.  Under CR conditions, one-dimensional string-like structures appear
($\langle z \rangle=2.04$, Fig.~\ref{fig:manybody}f), compared to folded
two-dimensional hexagonal packed monolayers with $\langle z \rangle=2.80$ that
form in the simulations employing CC conditions (Fig.~\ref{fig:manybody}e).

Arguably, open structures similar to Fig.~\ref{fig:manybody}f have been
observed for conducting particles in a low-permittivity
medium~\cite{barros14b}. However, we emphasize that the underlying mechanism
is different. In the case of dielectric mismatch, the total charge on each
individual particle is conserved and the polarization charge is redistributed
across the surface of the particle; the conductivity of the particles then
merely guarantees a constant potential on each surface. This differ from the
CR process, where the CP limit would be realized by globally grounding all
particles to a common potential, allowing free redistribution of charge among
different objects and the solution.  Such a system has, to our knowledge, not
been investigated.

\begin{figure}
\centering
  \includegraphics[width=\figurewidth]{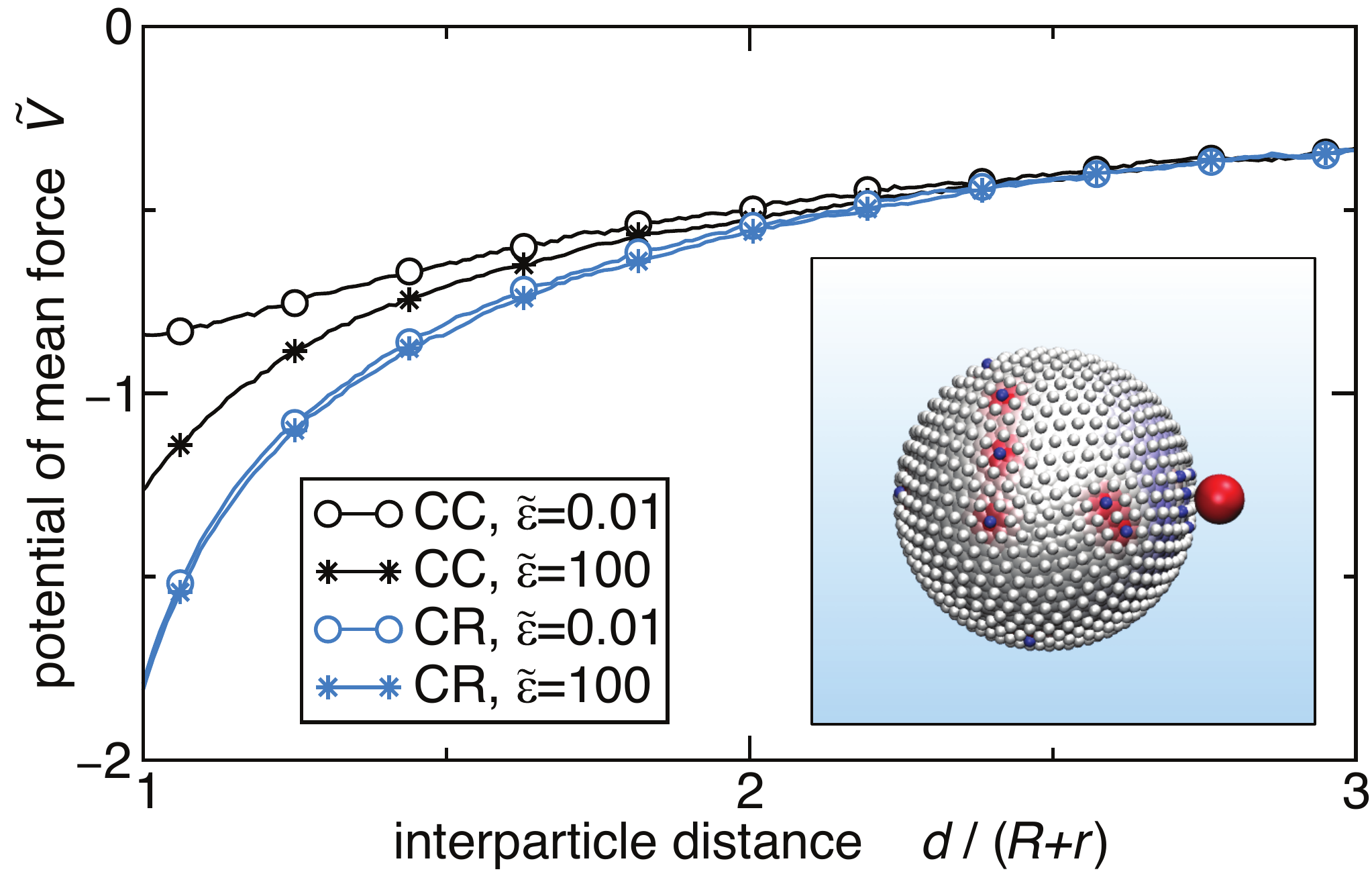}
  \caption{Dielectric effects on the (normalized) potential of mean
    force~$\tilde{V}$ for the particle pair described in Fig.~\ref{fig:pmf},
    with the additional condition that the large particle is polarizable.
    Under constant-charge (CC) conditions, a particle with low dielectric
    constant ($\tilde{\varepsilon}=0.01$) has a diminished attraction (open
    circles) and a high-dielectric particle ($\tilde{\varepsilon}=100$) has an
    enhanced attraction (asterisks). Both cases are superseded by the
    attraction strength under charge-regulating (CR) conditions. Here, the
    high-permittivity case still is slightly stronger, but the effect is
    barely visible on the scale of the graph.  The inset illustrates the
    CR-BEM model. The BEM layer of polarizable surface patches
    ($\tilde{\varepsilon}=100$) is placed at a distance~$R/8$ below the CR
    layer, with positive (red) or negative (blue) induced dielectric
    charges. The CR layer is shown as small white (neutral) or charged (blue)
    spheres.}
\label{fig:pmfdiele}
\end{figure}

Of particular interest, then, is the question of the combined effect of CR and
dielectric polarization. Like CR, polarization leads to charge redistribution
and accompanying strong many-body effects~\cite{ziwei20}. Moreover, the
prerequisite condition, namely a strong permittivity contrast between
particles and the surrounding medium, occurs in numerous aqueous systems,
including suspensions of silica or polystyrene colloids and protein solutions.
To answer whether CR or polarization dominates, we augment our particle model
with an additional boundary-element layer of 1472 patches uniformly
distributed on each sphere, positioned just below the CR layer in the inset of
Fig.~\ref{fig:pmfdiele}. Dielectric polarization charges are controlled by the
mismatch~$\tilde{\varepsilon}$, which denotes the ratio of the dielectric
constants of the particle and the surrounding solvent. After each MC and MD
step, we employ the IDS to compute the induced charge on each surface
element. Conversely, these polarization charges are taken into account when
computing the dissociation probability of each surface group. We evaluate the
role of dielectric effects by reexamining the system of Fig.~\ref{fig:pmf} for
two extreme cases: A small particle of fixed charge interacting with a large
particle of either high permittivity (i.e., nearly conducting;
$\tilde{\varepsilon}=100$) or low permittivity ($\tilde{\varepsilon}=0.01$). 
 As is well known~\cite{barros14b}, for CC boundary conditions the attraction
between the large and the small particle is suppressed in the low-permittivity
case and enhanced in the high-permittivity case (Fig.~\ref{fig:pmfdiele}).
Remarkably, however, CR almost completely screens any dielectric polarization
charges, yielding a PMF that is nearly independent of~$\tilde{\varepsilon}$
(cf.\ overlapping curves in Fig.~\ref{fig:pmfdiele}).  This observation is
consistent with Kirkwood's explanation of the dielectric increment of protein
solutions~\cite{kirkwood52}. Charge regulation screens dielectric
polarization, therefore, in the far field proteins can behave as high-dielectric
objects even though the protein core has a dielectric constant significantly
lower than water.

\begin{figure}
\centering
  \includegraphics[width=\figurewidth]{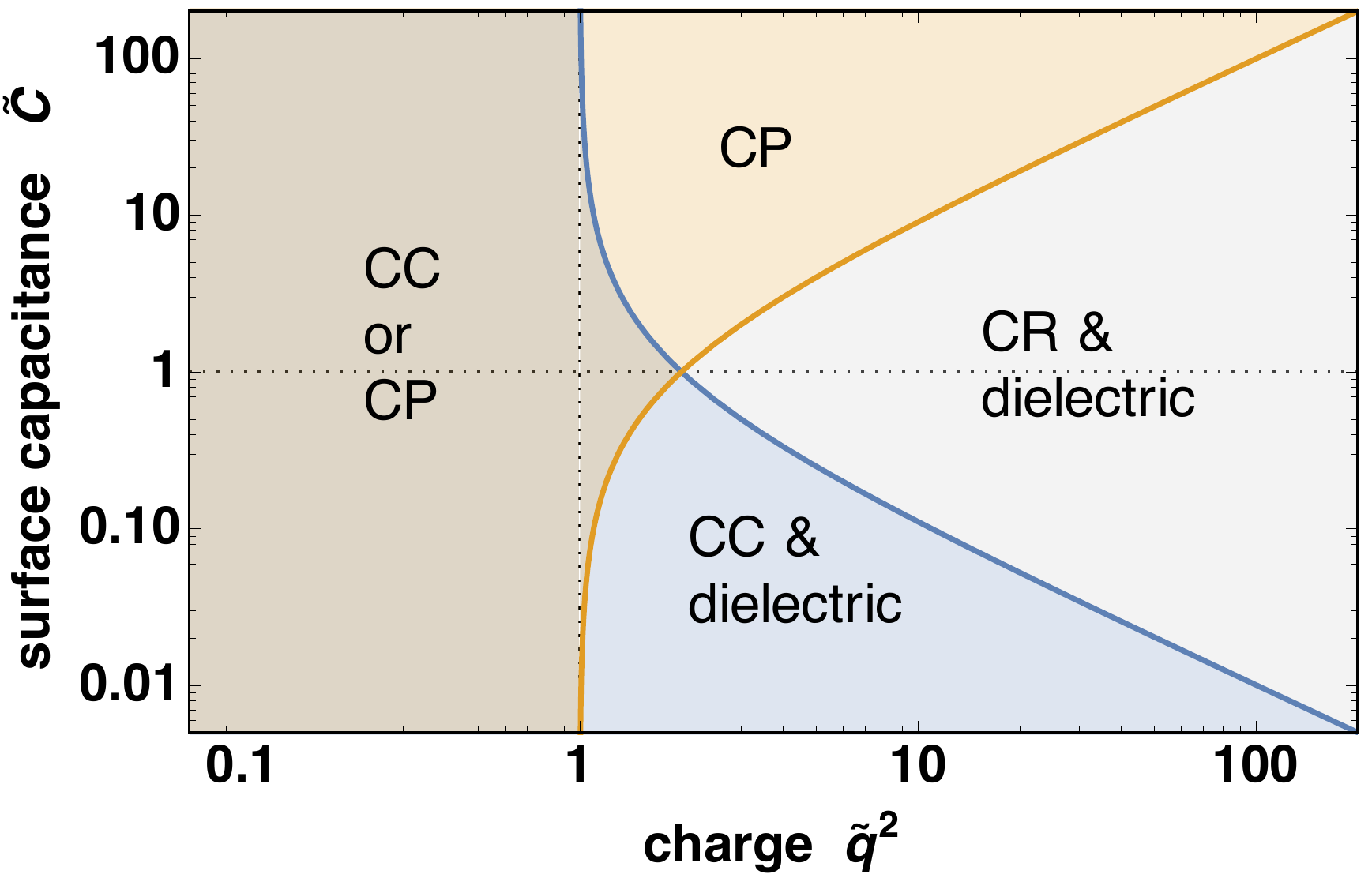}
  \caption{Schematic diagram showing the applicability of the constant-charge
    (CC) and constant-potential (CP) approximations of CR\@.  For sufficiently
    strong charges and intermediate values of the surface capacitance, the use
    of a full CR solver---as proposed in the main text---is required. In
    addition, in the two regions denoted by ``dielectric'' dielectric
    polarization effects can be important.}
\label{fig:schematic}
\end{figure}

In view of the potentially far-reaching consequences of CR, it is an important
question under which circumstances its effects can be ignored.
Figure~\ref{fig:pmfdiele} illustrates that for a small particle with
relatively high density of surface sites (a 6-nm silica particle in water), CR
strongly affects the interactions. Conversely, we can estimate when the CC or
CP approximation results in an error in the electrostatic interaction that is
smaller than the thermal energy~$k_{\mathrm{B}}T$.  We place a point
charge~$q$ at a distance~$d$ from a charge-regulating surface with a mean
surface charge density~$\sigma$ and a maximum (fully ionized) charge
density~$\sigma_0$.  In a mean-field approximation, $\alpha_i=\sigma/\sigma_0$
and the surface capacitance, determining the linear response of the surface
charge, then follows from Eq.~\eqref{eq:alpha} as
$C = \frac{\partial \sigma}{\partial \psi} = \sigma(1-\sigma/\sigma_0)(-\beta
q_0)$.  In the absence of ionic screening, the change in the potential at the
surface due to charge~$q$ is $ \Delta \psi = q/(4\pi \varepsilon d)$.  The
charge produced by the surface capacitance will be contained within an area of
size~$\sim d^2$ since $d$ is the relevant length scale. The additional charge
density due to CR is thus
$\sigma_{\mathrm{CR}} \sim - q \tilde{C}/[d^2(1+\tilde{C})]$ \footnote{The
  total potential change $\Delta \psi$ at the surface will be approximately
  $\beta \Delta \psi q_0^2 / l_{\mathrm{B}} \sim q/d - \sigma_{\mathrm{CR}}
  d$.}, with the dimensionless
capacitance~$\tilde{C}\equiv - Cd l_{\mathrm{B}}/(q_0^2 \beta)$. We observe
that the CC approximation is valid if the CR charge is sufficiently small,
$-\sigma_{\mathrm{CR}} q d \ll q_0^2/l_{\mathrm{B}}$ or
$\tilde{C} \ll 1/(\tilde{q}^2 - 1)$, where we define the reduced
charge~$\tilde{q} \equiv q/q_0 \sqrt{l_{\mathrm{B}}/d}$.  Conversely, the CP
limit implies an image charge~$q_{\mathrm{im}} \sim -q$, because for a single
flat surface the global CP limit is equal to the local CP condition and can
therefore be captured by a single image charge.  The CP limit is justified
when
$(\sigma_{\mathrm{CR}} d^2 - q_{\mathrm{im}}) q / d \ll q_0^2/l_{\mathrm{B}}$
or $ \tilde{C} \gg \tilde{q}^2 - 1$. The CP condition screens any possible
dielectric polarization charges and thus dielectric effects can be neglected.
These two conditions can be parametrized by just two dimensionless variables,
$\tilde{C}$ and~$\tilde{q}$, allowing us to delineate the different regimes in
Fig.~\ref{fig:schematic}.

This schematic also allows us to estimate the importance of CR for different
particle sizes.  Notably, the PMF in Fig.~\ref{fig:pmfdiele} was obtained for
a particle of only a few Bjerrum lengths in diameter ($R=4l_{\mathrm{B}}$),
corresponding to $\tilde{C} \sim 0.1$ and $\tilde{q}^2 \sim 200$. This
condition indeed falls inside the CR region in Fig.~\ref{fig:schematic}. Note
that the difference between high/low dielectric spheres in
Fig.~\ref{fig:pmfdiele} is very small, but still larger than $k_{\mathrm B}T$.
Rescaling the (linear) system size $d \to \gamma d$ while keeping the Coulomb
energy constant implies $q \to \gamma^{1/2} q$ and, therefore,
$\sigma \to \gamma^{-3/2} \sigma$. This rescaling keeps $\tilde{q}$ constant,
but the surface capacitance $\tilde{C}\sim -\sigma d l_{\mathrm{B}} / q_0$
changes as $\tilde{C} \to \gamma^{-1/2} \tilde{C}$~\footnote{Assuming low
  occupancy of surface sites, $\alpha \to 0$, which is expected for
  sufficiently large particles.}. Thus, the CC approximation becomes
increasingly more accurate as the particle size increases, which helps explain
why the CC approximation works rather well for predicting experimentally
observed crystal structures and clusters of micron-sized colloidal
particles~\cite{leunissen05,demirors15}.  On the other hand, nanoscale
particles will generally exhibit very strong CR effects; e.g., our results
(Fig.~\ref{fig:manybody}b) provide a possible explanation for the chain
formation observed in nanoparticle assembly~\cite{tang02,zhang15}.

In summary, we have implemented a hybrid MD--MC technique for resolving CR
effects in arbitrary dynamical systems. Utilizing this method, we have shown
that CR-induced many-body effects can qualitatively alter the predicted
self-assembled structures via stabilization of asymmetrically charged
aggregates. Both our numerical results and a scaling analysis demonstrate that
CR is particularly important for charged objects that are a few Bjerrum
lengths in size, such as proteins or nanoparticles, in which case CR screens
dielectric polarization effects.  Our method as well as the general findings
are broadly applicable to macromolecular and colloidal systems~\footnote{An
  open-source implementation in the LAMMPS MD package~\cite{plimpton95} is in
  preparation~\cite{charge_regulation_long}.}.
\nocite{plimpton95,charge_regulation_long}

Although we focused on acid/base dissociation, the method outlined in this
Letter can be directly utilized to study association or dissociation of
arbitrary ionic groups. For example, the MC part of our approach,
Eq.~\eqref{eq:alpha}, is equivalent to existing adsorption models for studying
calcium binding to proteins~\cite{Roosen-Runge2014}.

\begin{acknowledgments}
  This work was supported by the E.U.\@ Horizon 2020 program under the Marie
  Sk\l{}odowska-Curie fellowship No.~845032 and by the U.S. National Science
  Foundation through Grant No.\ DMR-1610796. We thank Jiaxing Yuan, Ziwei
  Wang, Gregor Trefalt, and Rudolf Podgornik for useful discussions and
  comments on the manuscript.
\end{acknowledgments}


%

\end{document}